\documentclass[twocolumns]{sdl2}

\usepackage[T1]{fontenc}
\usepackage{lmodern}

\def\0{\mbox{\tiny $0$}}
\def\1{\mbox{\tiny $1$}}
\def\2{\mbox{\tiny $2$}}
\def\3{\mbox{\tiny $3$}}
\def\4{\mbox{\tiny $4$}}
\def\5{\mbox{\tiny $5$}}
\def\6{\mbox{\tiny $6$}}
\def\7{\mbox{\tiny $7$}}
\def\8{\mbox{\tiny $8$}}
\def\9{\mbox{\tiny $9$}}

\usepackage{lipsum}
\usepackage{mathtools}
\usepackage{cuted}

\def\w{\mathrm{w}_{\0}}
\def\ws{\mathrm{w}_{\0}^{^2}}

\def\t{\frac{\hbar\,t}{m}}

\def\inc{_{_{\mathrm{INC}}}}
\def\re{_{_{\mathrm{REF}}}}
\def\tr{_{_{\mathrm{TRA}}}}

\newcommand{\inte}[1]{
\int
\mathrm{d} #1 \, \, 
}

\newcommand{\inted}[2]{
\iint
\mathrm{d} #1 \, \mathrm{d} #2 \, \, 
}

\newcommand{\intel}[3]{
\int_{_{#2}}^{^{#3}}\hspace*{-0.35cm} 
\mathrm{d} #1 \, \, 
}

\def\xs{x_{*}}

\def\zs{z_{*}}

\def\xt{\widetilde{x}}

\def\zt{\widetilde{z}}
\def\uxt{\tilde{x}}

\def\uzt{\tilde{z}}

\logo{
\colorbox{DarkGoldenrod}{\color{white}$\mathbf{\Sigma\hspace*{0.06cm} \delta \hspace*{0.04cm} \Lambda}$}
}

\journal{\textbf{\color{DarkRed} \normalsize
European Physical Journal Plus {\color{PrussianBlue}{137}}, 455-13 (2022)}}

\titlelines{2}
\title{From the delay time in quantum mechanics \\
to the Goos-H\"anchen shift in optics}

\imgbgabstract{
Delay times in quantum mechanics always represented an intriguing challenge for physicists. Due to the fact that
quantum mechanical  experiments  are, often, hard to be implemented, the possibility to connect delay times with 
laser lateral displacements gives us the opportunity to prepare, in optical laboratories, experiments which are \textit{equivalent} to the quantum mechanical ones in detecting delay times. In this article, we will show in detail not only the relationship between delay times and Goos-H\"anchen shifts, but also the close connection between the impulse change in quantum mechanics and angular deviations in optics. Lateral shifts  are caused by the  phase of Fresnel coefficients  whereas angular deviations by the  breaking of symmetry of the wave number distributions. The classical formula for the delay time is based on the use of the stationary phase method and contains a divergence  for incidence at a critical potential energy. For Gaussian beams, the mean value calculation removes such a divergence. The closed expression for the delay time for incidence at critical energy show an excellent agreement with the numerical calculation. The three-dimensional analysis of delay times allow to find the final and definitive connection between wave packets reflected by a potential in quantum mechanics and optical beams reflected by a dielectric/air interface.
}

\author{
\names{Stefano De Leo$^{1,a}$ and Leonardo Solidoro$^{2}$}
\affiliation{$^{1}$Institute of Mathematics, Statistics and Computing Science, Campinas State University, Brazil}
\affiliation{$^{2}$Salento University, Lecce, Italy}
\email{$^{a}$deleo@ime.unicamp.br}
}

\def\figureA{
\WideFigure{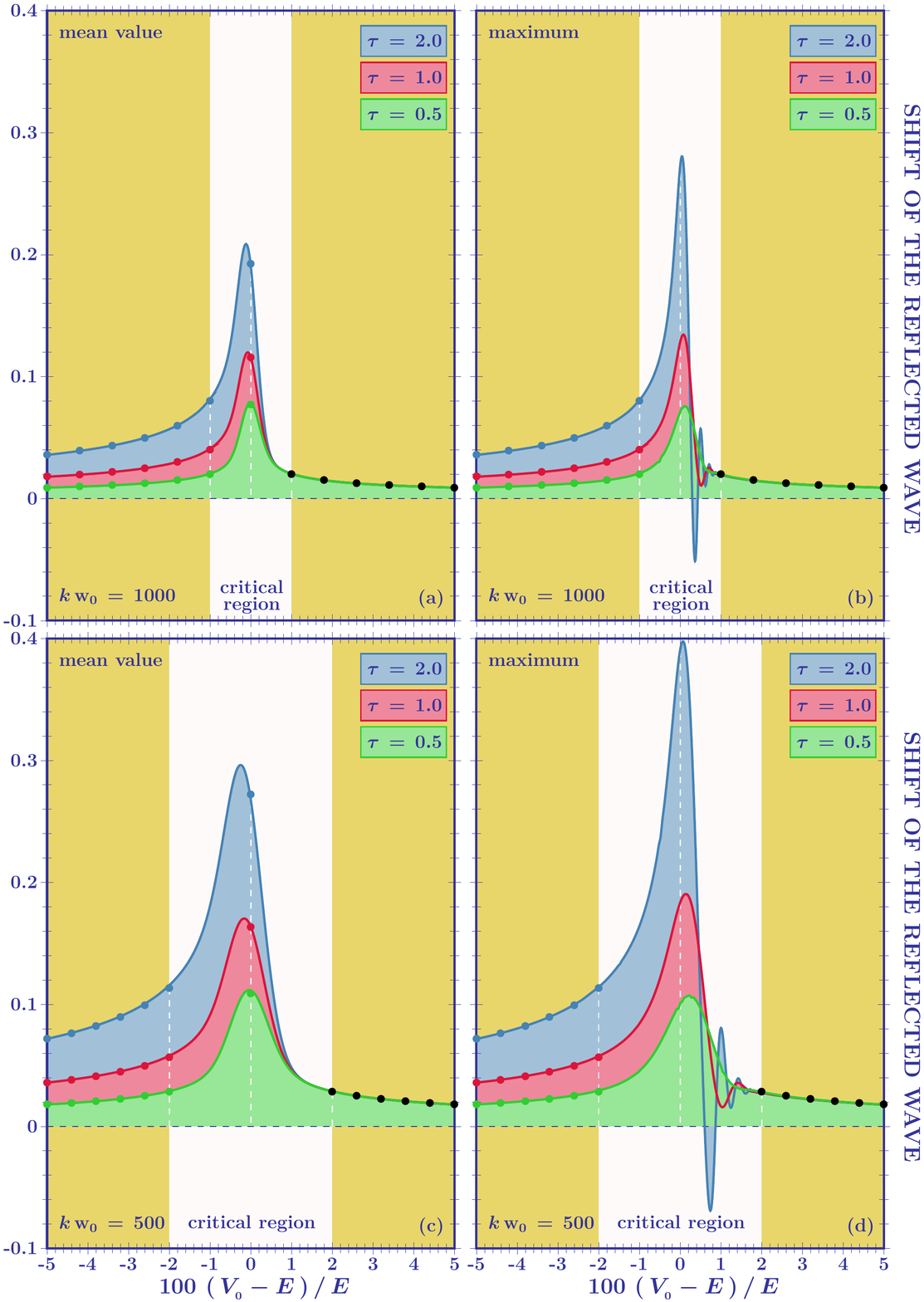}{Numerical simulations (\emph{continuous lines}) and analytical predictions (\emph{dots}) for the (adimensional) shift of the center of the reflected beam. The analytic results  are in excellent agreement with the numeric calculations outside the critical region (white color). At the left side of the critical region, we confirm the time dependence of the shift. This is an evidence of the changing in the modulus of momentum of the particle after reflection. At the right side, the numerical results confirm the analytical delay time. In this case the incident beam  is totally reflected with the same momentum (in modulus) of the incident one. In the critical region the mean value calculation (\ref{mvr}) appear as a dot and show agreement with the numerical curves. In such a region, it is also clear the Gaussian breaking of the symmetry of the reflected beam with respect to the incident one.\label{fig1}}{0.9}
}

\def\figureB{
\WideFigure{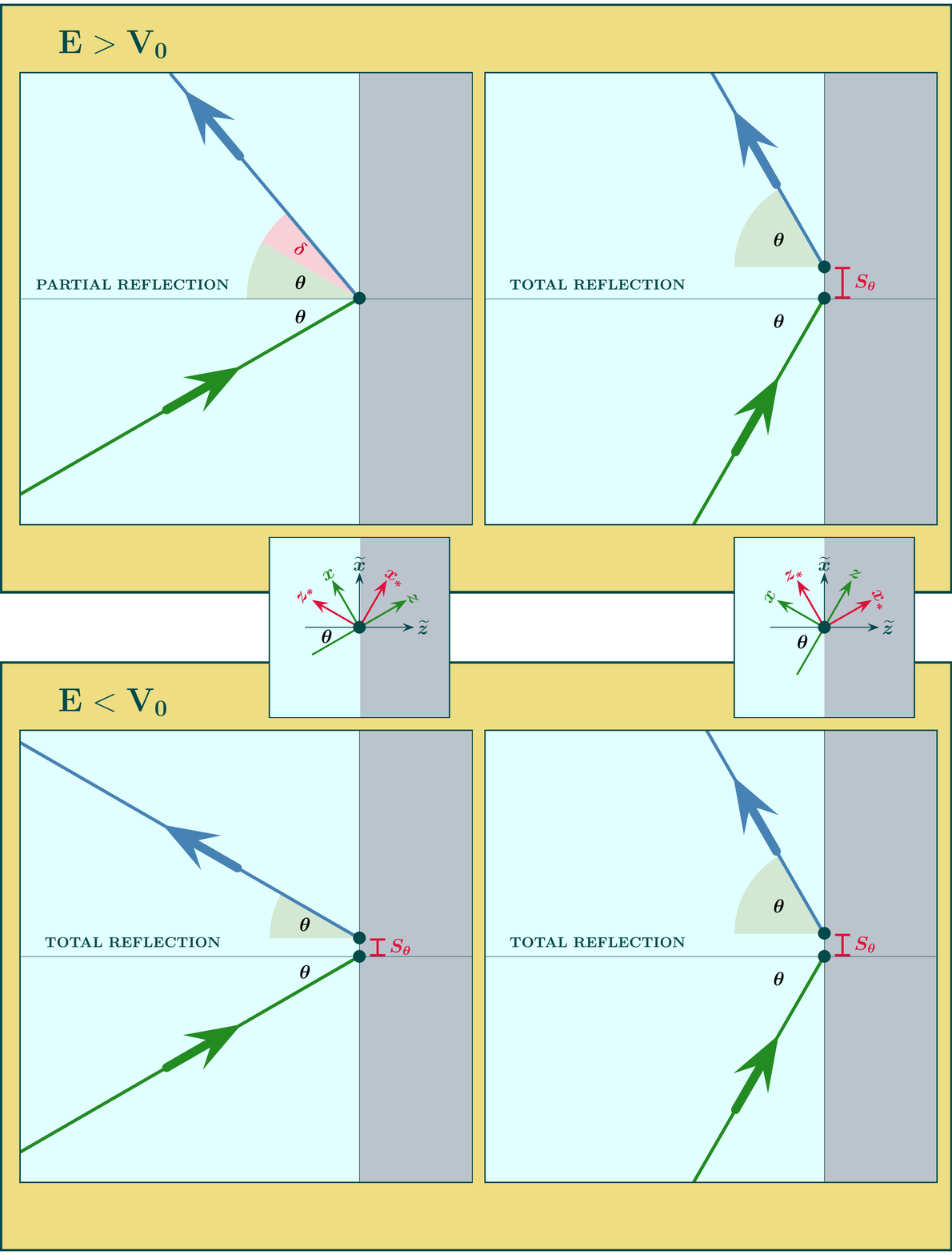}{Schematic representation of reflection for an electron beam with energy $E$ incident on a potential $V_{\0}$ in the plane $(x,z)$. For $E>V_{\0}$, we have partial reflection and angular deviations when the incidence angle is lower than the critical one (in the plots the critical angle is fixed at $\pi/4$) and we find total reflection and the Goos-H\"anchen shift for incidence angles greater than the critical ones. For $E<V_{\0}$, independently of the incidence angle, we always have the phenomenon of total reflection and of a lateral displacement. \label{fig2}}{0.9}
}

\def\figureC{
\WideFigure{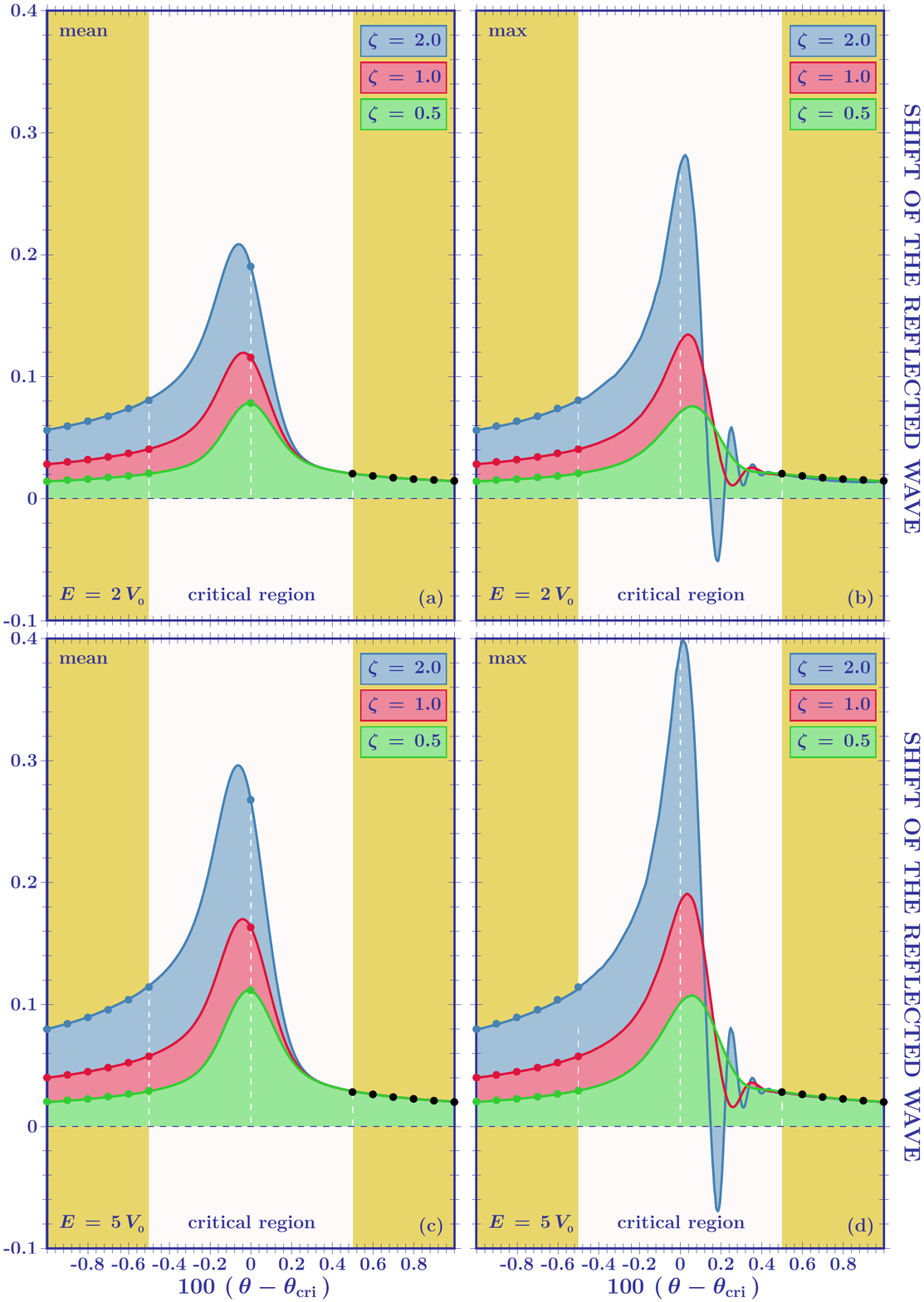}{Numerical simulations (\emph{continuous}) and analytical predictions (\emph{dots}) for an electron wave packet incident on a potential $V_{\0}$ with discontinuity along the $z$ axis. The plots refer to the mean value and maximum calculation. The energy and beam waist were fixed to $k\,\w=500$ (this characterizes the critical region which appear in white color), and the ratio between incidence energy and potential to 2 and 5 (which determine the critical angle). Three axial distance, $\zeta=0.5,\,1.0,\,2.0$, were examined. Numerical results show an excellent agreement with the analytical predictions. Outside the critical region the mean value and maximum calculation coincide and this is due to the symmetry of the beam. The shifts refer to $x_*/\w$. \label{fig3}}{0.9}
}

\def\figureD{
\WideFigure{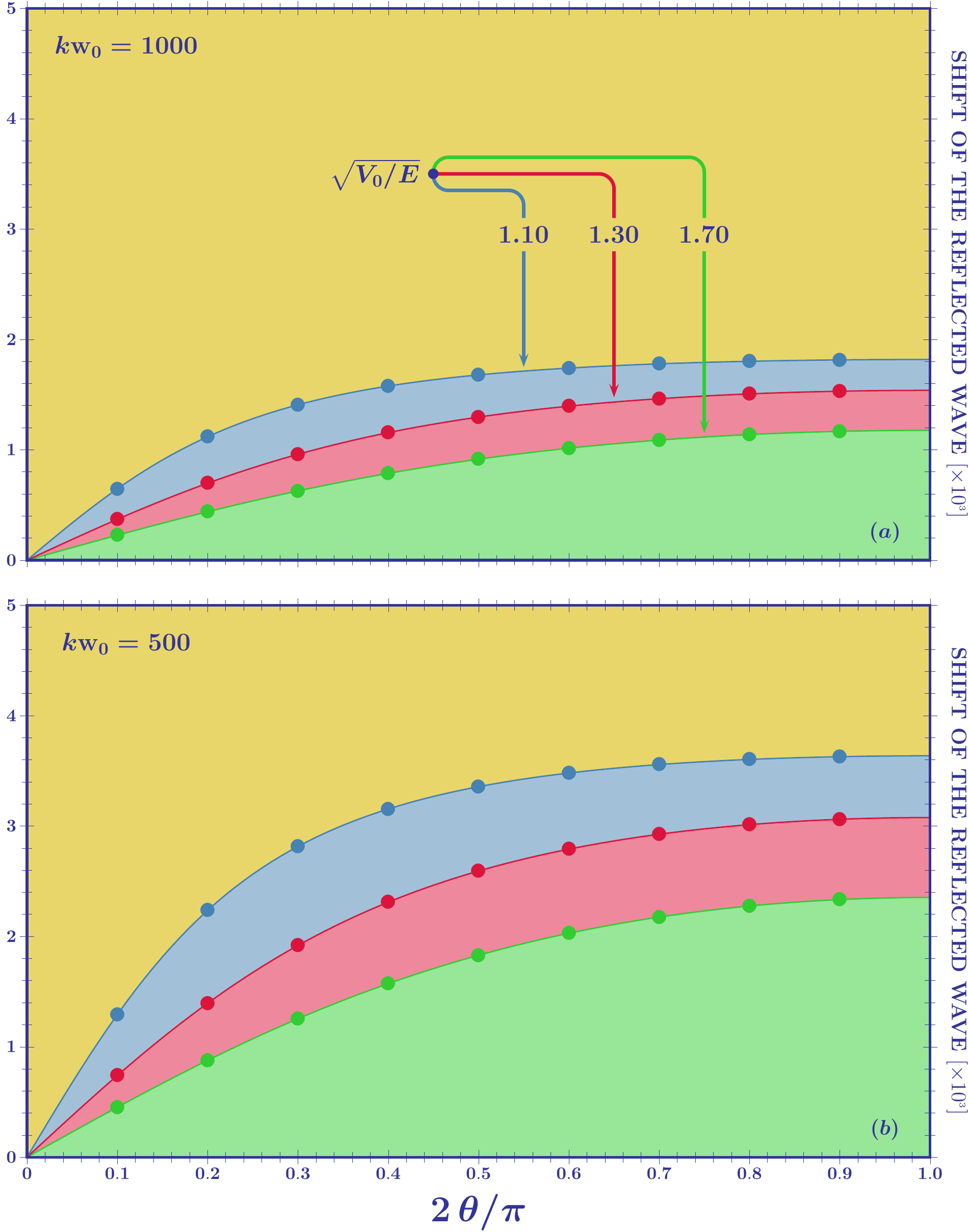}{Numerical simulations (\emph{continuous}) and analytical predictions (\emph{dots}) for the shift of the electron beam  in the case of  $E<V_{\0}$. Three values of $\sqrt{V_{\0}/E}$ were used, i.e. 1.10, 1.30, and 1.70.  In the limit $E \to V_{\0}$, the shift, $x_*/\w$, tends to $2/k\,\w$.
\label{fig4}}{0.9}
}

\def\gs{Goos-H\"anchen }

\begin{document}

\sdlmaketitle

\section{Introduction}

As \gs shift, we mean the lateral displacement of an optical beam totally reflected by a dielectric/air interface with respect to its incident point. The first experimental evidence of such a phenomenon appeared in literature 
in \cite{GH47}. In their experiment, the German physicists Goos and H\"anchen showed the evidence of such   \textit{anomalous} displacement of light  by using an elongated dielectric block which allowed multiple internal reflections to amplify the effect.  A theoretical explanation was given by Artmann  one year later \cite{AR48}. 
He proposed a mathematical description for the lateral displacement. By  assuming  that the plane waves which characterize the  electromagnetic field  have phases which rapidly vary, Artmann observed that the optical path is then  determined only by the main term of the phase, i.e. by the \emph{stationary condition}. In this framework, 
the \emph{shift} is then caused by the  additional phase which, in the case of total reflection, appears in the Fresnel coefficients. He also noted that due to the fact that transverse electric and magnetic waves have different expression for these coefficients, the lateral displacement depends on the beam polarization.   
The Artmann prediction was soon  confirmed by a new experiment of  Goos and H\"anchen \cite{GH49}.

Notwithstanding the success of the Artmann  formula, his prediction  contained a divergence for incidence at the critical angle. The divergence problem was addressed by the same physicist \cite{AR50} by analysing the case of  a  large number of internal reflections between two parallel surfaces. The research for a new formula became particularly prolific in the '60s  \cite{BR60,RE64,LO68} and '70s\cite{LD08}, with articles discussing 
the interval validity of the Artmann  formula, the connection between the shift and the energy conservation for totally reflected beams, and proposing new closed formulas. In 2016, \cite{AR16}, Ara\'ujo et al. obtained an  analytical formula based on the modified Bessel function which remove the the divergence and show an excellent agreement with the numerical calculation and recently also with experimental data \cite{LD19}.  A detailed description of the  Goos-H\"anchen \emph{shift} and  \emph{angular deviations} is found in \cite{DeL19}.

The Goos-H\"anchen shift has a quantum mechanical analogue when the first medium  is replaced by the vacuum and  the second one by a constant potential, and the optical field by a quantum-mechanical wave packet\cite{AnaQM}. In this case, the quantum particle move  in the $x/y$ plane and the 
discontinuity of the potential is along the $x$ axis, we shall cam back to this discussion later in section III. This allows an immediate analogy between the Goos-H\"anchen sfhit in Quantum Mechanics\cite{Cohen} and Optics\cite{Born}. Nevertheless, another interesting quantum phenomena such as  the delay time in the reflection by a potential step and/or  the tunneling time in the transmission  through a  potential barrier can be connected to optical phenomena.   In fact, the time-independent Schrödinger equation for quantum particles and the Helmholtz equation for electromagnetic waves are identical in form\cite{report}. Delay and tunneling times have to be calculated by analysing the peak of the reflected or transmitted wave packet and, so, can be translated in spatial shifts. This allows
an immediate connection with the Goos-H\"anchen optical shift.  In quantum mechanics, delay times in reflection and tunneling times in transmission lead to divergences that lead to infinite delay times in reflection from a potential barrier or to superluminal transmissions in the case of a portential barrier. We refer to the report cited in \cite{report} for a complete an clear revision of this intriguing phenomena. The divergences can be solved by using numerical calculation. These divergences are mathematically equivalent to those found, during the last years,  in Optics  when the 
Goos-H\"anchen  shift was discussed and the Artmann divergences removed\cite{AR16}. The main objective of this paper is to apply the optical technique used in Optics to see as the Taylor expansion of the reflected wave packet  distribution can be useful to remove the divergences found in Quantum Mechanics. If the delay time, by analysing the displacement  of the reflected  peak, can be linked to the Goos-Hanchen shift of an optical beam, an immediate question arises: which is the mechanical counterpart of angular deviations? Angular deviations in Optics\cite{DeL19} are  caused by the symmetry breaking of the angular distribution of the optical beam \cite{DeL19}, so, in Quantum Mechanics, we expect that a similar symmetry breaking occurs in the wave number distribution leading to a change in the group velocity of the reflected packet. In the next Section, we will analyse in detail two cases: the one of incidence with energy higher than the potential step  which leads to a wave number distribution symmetry breaking and the consequent change in the group velocity of the reflected particle and the one of incidence with lower energy which leads to delay times but not at the speed change of the reflected packet. In our analysis, we also differentiate between peak and mean calculation. The results obtained by the  analytical study based on the Taylor expansion of the wave number distribution of the reflected beam found an excellent agreement with numerical simulations. Finally,  the adimnesional analysis also allows to extend our results to different experimental optical setups and this is important in view of possible optical simulations of quantum mechanical problems. The didactic presentation aims to reach a wide audience and stimulate further discussions.

\section{Velocity change and delay time}
The dynamics of a non-relativistic particle with mass $m$ and  moving along the $x$-axis is described, in the presence of a potential, by a wave function satisfying the one-dimensional Scrh\"odinger equation\cite{SC,Cohen}: 
\begin{equation}
\label{eq:scr}
i\,\hbar\,\Psi_{t}(x,t)\,=\,-\,
\frac{\hbar^{^2}}{2\,m}\,\Psi_{xx}(x,t)\, +\, V(x)\,\Psi(x,t) \,\, ,
\end{equation}
where $V(x)$ is null for $x<0$ and has a constant value $V_{\0}$ for $x>0$. For a particle with energy $E$, we choose, for the incident wave packet $(x<0)$,  a Gaussian distribution centred on $k=\sqrt{2\,m\,E}/\hbar$,
\begin{align}
\label{free}
\Psi\inc(x,t)&=\inte{k_x} g(k_x-k) \nonumber\\
&\times\,\, \exp\left[i\left(\,k_x\,x\,-\,\frac{k_x^{^{2}}}{2}\, \t \,\right)\right]\,\,,
\end{align}
where
\begin{equation*}
 g(k_x - k)\,=\,\w\,\exp\,\left[\,-\,(\,k_x\,-\,k\,)^{^{2}}\w^{\,\2}\,/\,4
\,\right]\,/\,2\,\sqrt{\pi}\,\,.    
\end{equation*}
After reaching the potential region at $x=0$, the incident beam splits in two beams, the reflected and transmitted one. The reflected beam, moving back in the region  $x<0$, is described by
\begin{align}
\label{ref}
\Psi\re(x,t)&=\inte{k_x}
R(k_x)\,g(k_x-k) \nonumber \\
 &\times\,\, \exp\,\left[-i\left(k_x\,x\,+\,\frac{k_{x}^{^{2}}}{2}\, 
\t \right)\right]\,\,,
\end{align}
and the transmitted one moving forward in region $x>0$ by
\begin{align}
\label{tra}
\Psi\tr(x,t)&=\inte{k_x}T(k_x)g(k_x-k) \nonumber \\
\times&  \exp\,\left[i\,
\left(\sqrt{k_x^{^2}-k_{\0}^{^{2}}}\,x-\frac{k_x^{^{2}}}{2}\, \t \right)\right]\,\,,
\end{align}
where $k_{\0}=\sqrt{2\,m\,V_{\0}}/\hbar$,  
\begin{equation}
R(k_x)\,=\,\frac{k_x\,-\,\sqrt{k_x^{^2}-k_{\0}^{^{2}}}}{k_x\,
+\,\sqrt{k_x^{^2}-k_{\0}^{^{2}}}}\,\,,
\end{equation}
and
\begin{equation}
T(k_x)\,=\,\frac{2\,k_x}{k_x\,
+\,\sqrt{k_x^{^2}-k_{\0}^{^{2}}}}\,\,.
\end{equation}
The reflection and transmission coefficients were obtained by imposing the continuity of the fields
$\Psi\inc(x,t)\,+\,\Psi\re(x,t)$ and $\Psi\tr(x,t)$ and their derivatives at $x=0$. 

The choice of the Gaussian distribution allows to give an analytical expression for the incident wave packet,
\begin{align}
\label{Free_ana}
 \MoveEqLeft[3] \Psi\inc(x,t)=\frac{\exp\left[i\left(k\,x - E\,t/\hbar\right)\right] }{\sqrt{1+2 \, i\, \,\tau}} \nonumber \\
  \times{}&\,
\exp\left[\,-\,\frac{\left(x - k \,\ws\,\tau\right)^{\2}}{\ws\,(1+2\,i\,\tau)}\right]\,\,,
\end{align}
where $\tau=\hbar\,t/ m\,\ws$ is the \textit{adimensional} time.  An analytically integration for the reflected wave packet is not possible due to the presence of $R(k_x)$ in the integrand. Nevertheless, the Gaussian distribution, $g(k_x - k)$, quickly decreases around its peak, so, we can  obtain an analytical expression for the reflected wave packet  by using the first order Taylor expansion of the reflection coefficient, i.e.
\begin{equation}
\label{approx}
R(k_x) \,\,\approx\,\, R(k) \,\left[\,1\,-\, \frac{2}{\sqrt{k^{^2}-k_{\0}^{^{2}}}}\,\,(k_x-k)\,\right]\,\,.
\end{equation}
By using this Taylor expansion and observing that the linear term, $k_x-k$, in the integrand can be substituted
a partial derivative in $x$, we can rewritten the reflected wave packets in terms of the incident one as follows 
\begin{multline}
\label{RefA}
\Psi\re(x,\tau)  =\frac{\exp\left[-\,i\left(k \,  x \, + \, E\,t/\hbar\right)\right]  }{ \sqrt{1+2 \, i \, \tau} }\,R(k) \\
\times\left(1-\frac{2\,i\,\partial_x}{\sqrt{k^{^2}-k_{\0}^{^{2}}}}\,\right)
\exp\left[\,-\,\frac{\left(x + k \,\ws\,\tau\right)^{\2}}{\ws\,(1+2\,i\,\tau)}\right]\,\,.
\end{multline}
After simple algebraic manipulations, we find
\begin{equation}
	\Psi\re(x,\tau)\,=\,\mathcal{R}(x,\tau)\,\Psi\inc(-x,\tau)\,\,,
\end{equation}
where
\begin{equation}
\label{fr}
 \frac{\mathcal{R}(x,\tau)}{R(k)}\,=\,\left[1+\frac{4\,i}{\ws\,\sqrt{k^{^2}-k_{\0}^{^{2}}}}\frac{x + k\,\ws\,\tau}{1+2 \, i \, \tau}\right]\,.
\end{equation}
At this point, we have to  distinguish between  two cases: Incidence energy above the potential  energy  ($k>k_{\0}$) and incidence energy below ($k<k_{\0}$).

\subsection{Beam centre for $\boldsymbol{k>k_{\0}}$}

For a non-relativistic particle with energy greater than the potential one, the square root which appear in the denominator of the reflection factor (\ref{fr}) is real and, consequently,  the intensity of the reflected beam is given by
\begin{align}
\label{IRR1}
\left|\, \Psi\re(x,\tau;k>k_{\0}) \, \right|^{^2}
=  R^{^2}(k) \frac{|\Psi\inc(-\,x,t)|^{^{2}}}{1+4 \,\tau^{\2}} \nonumber \\
 \times \left\{1+ \left[\, 2\,\tau\,+\,\frac{4\left(x + k\,\ws\,\tau\right)}{\ws\,\sqrt{k^{^2}-k_{\0}^{^{2}}}}
\,  \,\right]^{^2} \right\}\,\,.
\end{align}
This expression can be further simplified by observing that due to the fact that  $|\Psi\inc(-\,x,t)|$ is centred 
at  $x =- k\,\ws\,\tau$  we can consider the first order expansion  of the quadratic factor which appears in the reflected intensity,
\begin{align}
\label{IRR2}
\left|\, \Psi\re(x,\tau;k>k_{\0}) \, \right|^{^2}
 \approx  R^{^2}(k) \,|\Psi\inc(-\,x,t)|^{^{2}} \nonumber \\
 \times\left[1\,+\,\frac{16\,\,\tau}{\ws\,\sqrt{k^{^2}-k_{\0}^{^{2}}}}
\, \frac{x + k\,\ws\, \tau}{1+4 \,\tau^{\2}}\,\right]\,.
\end{align}
It is interesting to observe that this expression 
also represents the first-order Taylor expansion of a Gaussian function
\begin{multline}
\label{IRR3}
\left|\, \Psi\re(x,\tau;k>k_{\0}) \, \right|^{^2}
=  \frac{R^{^2}(k)}{\sqrt{1+4 \, \tau^{\2}}} \\
 \times \exp\left[-\,2\,\frac{\left(\,x + k\,\ws\, \tau - \,4\,\tau/\sqrt{k^{^2}-k_{\0}^{^{2}}}\right)^{\2}}{\ws\,\left(1+4\,
\tau^{\2}\right)}\right]\,.
\end{multline}
The Gaussian approximation  allows to immediately  obtain the centre of the reflected beam, 
\begin{equation}
\label{shift1}
x\re^{^\mathrm{centre}}(k>k_{\0})\,=\,-\, k\,\ws\,\, \tau \,+\,\frac{4}{\sqrt{k^{^2}-k_{\0}^{^{2}}}}\,\tau\,\,.
\end{equation}
The formula shows  a time dependent  shift of the center of the reflected beam with respect to the position  $-\, k\,\ws\,\, \tau$ expected for a particle incident with wave number  $k$ and reflected by a potential with wave number  $k_{\0}<k$.  The time dependence implies a change in the group velocity, $\hbar\,\widetilde{k}/m$,  of the reflected wave 
\begin{equation}
\widetilde{k}\,=\, k\,-\, \frac{4}{\sqrt{k^{^2}-k_{\0}^{^{2}}}\,\,\ws}\,\,.
\end{equation}
The ratio $\widetilde{k}/k$  depends not only on the incidence energy and  the potential height but also on 
the waist, $\w$, of the incident beam. This interesting result means that a measure of the momentum of the reflected beam can be used to obtain information on the beam parameter $\w$.    The ratio between the group velocity of reflected and incident beams is, finally,  given by
\[
\frac{\widetilde{v}}{v}\,=\,1\, -\, \frac{4}{\sqrt{k^{^2}-k_{\0}^{^{2}}}\,\,k\,\ws}\,\,.
\]   
For an incident energy  $E=1\,\mathrm{eV}$ (this implies an electron velocity of $2\,c/10^{^{3}}$) and a beam waist $\w=0.2\,\mu\mathrm{m}$, we have $k\,\w=10^{^{3}}$. Consequently, for $E\gg V_{\0}$,  the difference between the incident and the reflected  velocity is of the order of $10^{^{-6}}\,v$. For an incident energy very close to the potential one,
\[ k\,\w\,=\,k_{\0}\,\w\,+\,\delta\,/\,k\,\w\,\,,    \] 
we find
\[
\frac{\widetilde{v}}{v}\,\approx\,1\, -\, \frac{4}{\sqrt{2\,\delta}\,\,k\,\w}\,\,,
\]   
which represent  an amplification of $10^{^{3}}$ with respect to the previous case. The divergence problem foun for incidence $k-k\0$ will be discussed later.  

\subsection{Beam centre for  $\boldsymbol{k<k_{\0}}$}
For $k<k_{\0}$, the reflection coefficient, $R(k)$, is a complex phase and we have total reflection, $|R(k)|=1$. Observing that  the square root in the denominator of $\mathcal{R}(k)$ in Eq.\,(\ref{fr})  is now responsible for an additional imaginary units, we find, for the reflected beam,   the following intensity 
\begin{align}
\label{IR1}
\left|\, \Psi\re(x,\tau;k<k_{\0}) \, \right|^{^2} = \frac{|\Psi\inc(-\,x,t)|^{^{2}}}{1+4 \,\tau^{\2}} \nonumber \\
\times\left\{\left[\, 1+\,\frac{4\left(x + k \,\ws\tau\right)}{\ws\,\sqrt{k_{\0}^{^2}-k^{^{2}}}}
\right]^{\2}\,+\, 4\,\tau^{\2} \right\}\, 
\,\,.
\end{align}
By neglecting the $x + k \,\ws\tau$ second order term in the Taylor expansion, we obtain
\begin{align}
\label{IR2}
\left|\, \Psi\re(x,\tau;k<k_{\0}) \, \right|^{^2}
 \approx   |\Psi\inc(-\,x,t)|^{^{2}} \nonumber \\
 \times \left[1\,+\,\frac{8}{\sqrt{k_{\0}^{^2}-k^{^{2}}}}
\, \frac{x + k\,\ws \tau}{\ws\,\left(1+4 \,\tau^{\2}\right)}\,\right]\,\,,
\end{align}
which, after some algebraic manipulations,  can be rewritten, as done in the previous subsection, in a Gaussian form, 
\begin{align}
\label{IR3}
&\left|\, \Psi\re(x,t;k<k_{\0}) \, \right|^{^2}
=   \frac{1}{\sqrt{1+4 \,\tau^{\2}}}\nonumber \\
&\times\,\exp\left[-\,2\,\frac{\left(x + k\,\ws\tau - 2\,/\,\sqrt{k_{\0}^{^2}-k^{^{2}}} \right)^{\2} }{\ws\,\left(1+4\,
\tau^{^2}\right)}\right]
\,\,.
\end{align}
The centre of the reflected beam is now found at 
\begin{equation}
\label{shift2}
x\re^{^\mathrm{centre}}(k<k_{\0})\,=\,-\,k\,\ws \tau \,+\,\frac{2}{\sqrt{k_{\0}^{^2}-k^{^{2}}}}\,\,.
\end{equation}
Observe that, in this case,the shift is  time independent. This means that the modulus of the velocity of the reflected particle is the \textit{same} of the incident one. This shift leads to the well-known phenomenon of \textit{delay time} \cite{Cohen}. Indeed by
\begin{equation*}
x\re^{^\mathrm{centre}}(k<k_{\0})\,=\,-\,k\,\ws (\,\tau\,-\,\tau_{\0}\,)\,\,,
\end{equation*}
we obtain
\begin{equation}
  t_{\0}= \frac{2\,m}{\hbar\,k\, \sqrt{k_{\0}^{^2}-k^{^{2}}}} = \frac{\hbar}{\sqrt{E\,(V_{\0}-E)}}\,\,.
\end{equation}

\subsection{Mean value calculation for $\boldsymbol{k=k_{\0}}$}

The divergence at $k=k_{\0}$ of Eqs.\,(\ref{shift1}) and (\ref{shift2}) can be removed by a mean value calculation,
\[
\frac{\displaystyle{\inte{x}}\,\,x\,  \left|\, \Psi\re(x,\tau;k=k_{\0}) \, \right|^{^2}}{\displaystyle{\inte{x}}\left|\, \Psi\re(x,\tau;k=k_{\0}) \, 
\right|^{^2}}\,\,.  \]
From $x$ integration we get a Dirac delta function which then  allows to further reduce the integration to a $k_x$ integration, 
\begin{equation}
\label{mean}
\left\langle \frac{x}{\w}\,+\,k_{\0}\,\w\, \tau \right\rangle\re = \frac{N_{_{\mathrm{A}}}+N_{_{\mathrm{B}}}}{D{\mbox{\footnotesize en}}}\,\,,
\end{equation}
where
\begin{eqnarray}
N_{_{\mathrm{A}}} & = & -\,\tau\,\ws\inte{k_x}(\,k_x-k_{\0}\,)|R(k_x)|^{^2}\,g^{\2}(k_x-k_{\0})\,\,\nonumber\\
N_{_{\mathrm{B}}}& = & 2\,\intel{k_x}{-\,\infty}{k_{\0}}\frac{g^{\2}(k_x-k_{\0})}{\sqrt{k_{\0}^{^2} - k_{x}^{^2}}}\,\,,\\
D{\mbox{\footnotesize en}}&=&\w\,\inte{k_x} |R(k_x)|^{^2}\,g^{\2}(k_x-k_{\0})\,\,. \nonumber
\end{eqnarray}
In order to obtain an analytical integration of the previous terms, let us introduce the  adimensional quantity
\[\alpha=(k_x-k_{\0})\,\w\,\,\]
and expand  the reflection coefficient, around $k_x\ge k_{\0}$, 
\begin{equation}
R(k_x)\,
\approx\,  1\,-\,2\,\sqrt{\frac{2\,\alpha}{\w k_{\0}}}\,\,.
\end{equation}
By using the adimensional integral parameter $\alpha$ and the reflection coefficient expansion, we can rewrite 
Eq.(\ref{mean}) as follow 
\begin{align}
\label{mean2}
\left\langle \frac{x}{\w}\,+\,k_{\0}\,\w\, \tau \right\rangle\re
  \,\,=\,\, \frac{1}{\sqrt{\pi\,k_{\0}\w}}\,\,\,\times \nonumber  \\
 \intel{\alpha}{0}{+\,\infty}\exp\left[-\,\frac{\alpha^{\2}}{2}\right] 
 \left(\,4\,\alpha^{\3/\2}\,\tau\,+\, \alpha^{-\,\1/\2} \,\right)\,\,.
 \end{align}
Finally the mean calculation removes the divergence at 
 \begin{equation}
 \label{mvr}
 \left\langle \frac{x}{\w}\,+\,k_{\0}\,\w\, \tau \right\rangle\re
 \,= \,  \frac{2^{^{\5/\4}}\,\Gamma\left(5/4\right)}{\sqrt{\pi\,k_{\0}\,\w}}\,\,(\,2\,\tau\,+\,1\,)\,\,.
\end{equation}

\subsection{Numerical analysis}

The analytic results obtained in the previous subsections have been  tested by numerical simulations. 
The expression found for the shift of the reflected wave, Eqs.\,(\ref{shift1}) and (\ref{shift2}), were obtained by using a Taylor expansion of the reflection coefficient $R(k_x)$, see Eq.\,(\ref{approx}).  This means that we expect 
for
\[
\frac{V_0-E}{E}\,<\, -\,\frac{10}{k\,\w}\,\,,\]
a time dependent shift the given by
\begin{equation}
\label{s1f1}
\frac{x\re^{^\mathrm{centre}}}{\w}\,+\,k\,\w \tau \,=\,\frac{4}{\sqrt{k^{^2}-k_{\0}^{^{2}}}}\,\tau\,\,,
\end{equation}
and for 
\[
\frac{V_0-E}{E}\,>\, \frac{10}{k\,\w}\,\,,\]
a delay time in the reflection given by
\begin{equation}
\label{s2f1}
\frac{x\re^{^\mathrm{centre}}}{\w}\,+\,k\,\w \tau \,=\,\frac{2}{\w\,\sqrt{k_{\0}^{^2}-k^{^{2}}}}\,\,.
\end{equation}
For $E=V_{\0}$, we shall use the mean value result obtained in Eq.\,(\ref{mvr}).

For two values of $k\,\w$, i.e. 500 and 1000, we numerically calculated the mean value and the maximum of the reflected beam. The plots (continuous lines) appear in Figure 1.  The use of two different values for $k\,\w$ confirms our prediction about the validity region  of the analytical results (dots).  In such regions, analytic and numeric results show a perfect agreement. The mean value and maximum calculations confirm that in the analytic region, the symmetry of the incident beam is conserved after reflection and so we can talk of center of the beam. Such a symmetry is broken in the critical region, see the white regions in Figure 1.  To test the time dependence of Eq.\,(\ref{s1f1}), the numerical plots appear in Figure 1 for  three temporal values, i.e. $\tau=0.5,\,1,\,2.$ 
At the left side of each panel the time dependence is clear and the plots show  the  change in the modulus of the mean momentum of the reflected beam with respect to the incident one.  At the right side, we find the shift caused by the delay time.   In this case, the modulus of the  mean momentum of the reflected beam is the same of the incident particle and
we have total reflection.

\section{Three-dimensional analysis}
In view of possible analogies between quantum mechanics and optics, let us consider Gaussian wave packets propagating in $z$ direction. The wave number distribution
\begin{equation}
G(k_x,k_y)\, = \,\ws\, \exp[\,-\,(\,k_x^{^2} + k_y^{^2}\,)\,\ws\,/\,4\,]\,/\,4\,\pi
\end{equation}
allows to describe the propagation of a free particle in the three-dimensional space, 
\begin{equation}
\label{3dg}
\Psi\inc(x,y,z;t) \,=\,\psi\inc(x,y,z)\,\exp[\,-\,i\,\frac{E\,t}{\hbar}\,]\,\,,
\end{equation}
where
\begin{align}
\label{3dgauss}
\psi\inc(x,y,z) &=\iint \mathrm{d}k_x\mathrm{d}k_y  \, G(k_x,k_y) \nonumber\\
&\times\exp\left[\,i\,(\,k_x\,x \,+\, k_y\, y \,+\, k_z \,z\,)\,\right]
\end{align}
with $k_z=\sqrt{k^{^{2}}-k_x^{^{2}}-k_y^{^{2}}}$ and   $k=\sqrt{2\,m\,E}\,/\,\hbar$.  The wave packet (\ref{3dg}) satisfies the free three-dimensional Schr\"odinger equation, 
\begin{equation}
i\,\hbar\,\partial_t\,\Psi\inc(x,y,z;t)\,=\,-\frac{\hbar^{^{2}}\nabla^{^{2}}}{2\,m}\Psi\inc(x,y,z;t) \,.
\end{equation}
In the case in which
\begin{equation}
\label{apx}
k_z \, \approx \, k - \frac{k_x^2 + k_y^2}{2\,k}\,\,,
\end{equation}
the integrals in Eq.\eqref{3dgauss} can be analytically solved, leading to
\begin{align}
\label{free_ana3d}
\psi\inc(x,y,z) &= \frac{\exp[\,i\,k\,z\,]}{1+2 \, i \, \zeta} \nonumber \\
&\times\exp\left[-\,\frac{x^{\2} + y^{\2}}{\ws\,(\,1+2\,i\,\zeta\,)}\right] \, \,,
\end{align}
where $\zeta=z/k\,\ws$.
We observe that, with respect to Eq.\eqref{Free_ana}, the spatial variable $\zeta$ takes the place of the time variable $\tau$.

As done in the previous section, we shall analyse the beam reflected by the  potential 
\begin{equation}
V(\xt,y,\zt)\,=\,\left\{\,
0\,\,\,\mathrm{if}\,\zt<0\,\,\,\,\,\mathrm{and}\,\,\,\,\,
V_{\0}\,\,\,\mathrm{if}\,\zt>0\,\right\} \, \,, 
\end{equation}
see Figure 2. The coordinate system $(x,\,y ,\,z)$ is the proper system of  the incident wave, while \emph{tilde} and \emph{starred} systems respectively refer   to the potential and reflected beam systems, i.e. $\zt$ is the direction perpendicular to the air/potential interface and the $\zs$ indicates the propagation axis of the reflected beam. 

Let $\theta$ be the angle of incidence. The relations between different coordinate systems is given by
\begin{align}
\label{cordRot}
&\begin{bmatrix}
\xt \\ 
\zt \\
\end{bmatrix}
\, = \, 
\begin{pmatrix}
\cos\theta & \sin\theta \\
-\sin\theta & \cos\theta \\
\end{pmatrix}
\begin{bmatrix}
x \\
z 
\end{bmatrix}
\quad
\mathrm{,} \nonumber \\
&\begin{bmatrix}
\zs \\ 
\xs \\
\end{bmatrix}
\, = \, 
\begin{pmatrix}
\sin\theta & -\cos\theta \\
\cos\theta & \sin\theta \\
\end{pmatrix}
\begin{bmatrix}
x \\
z 
\end{bmatrix}\,.
\end{align}
The same relations hold for wave-vector components.
The reflection coefficient can be derived directly from continuity conditions of the field and its derivative,
\begin{equation}
\label{refcoef2d}
R(k_x,k_y) = \frac{k_{\uzt}-q_{\uzt}}{k_{\uzt}+q_{\uzt} }\,,
\end{equation}
where
\[
k_{\uzt} = \sqrt{k^{^2}-k_{\uxt}^{^2}-k_{y}^{^2}} \,\,\,\, \mathrm{and}\,\,\,\, 
q_{\uzt} = \sqrt{q^{^2}-k_{\uxt}^{^2}-k_{y}^{^2}}\,,
\]
and    $\hbar\,q = \sqrt{2\,m\,(E-V_0)}$. It is interesting to observe that the condition 
$k_{\uxt}=q_{\uxt}$ is due to the fact that the potential changes along the $\zt$ direction. Such a condition 
implies 
\begin{equation}
\label{kqsnell}
k\sin\theta\,=\,q\sin\varphi\,,
\end{equation}
where $\varphi$ is the angle that the transmitted beam forms with $\zt$. The previous equation he be seen as   
the electron counterpart of the Snell law (we shall come back later on this point).

Let us know calculate the integral form of the reflected beam. By a $\theta$ rotation, the spatial phase of the incident beam can be expressed in terms of the proper axes of the potential, i.e.
$k_{\uxt}\,\xt + k_y\,y + k_{\uzt}\,\zt$. The spatial phase of the reflected beam can be rewritten as follows    
\begin{eqnarray}
k_{\uxt}\,\xt - k_{\uzt}\,\zt & = &
\begin{bmatrix}
k_{\uxt} &  k_{\uzt}
\end{bmatrix}
\begin{pmatrix}
1 & 0 \\
0 & -1 
\end{pmatrix}
\begin{bmatrix}
\xt \\
\zt 
\end{bmatrix}\nonumber\\
 & = & 
\begin{bmatrix}
k_x &  k_z
\end{bmatrix}
\begin{pmatrix}
0 & 1 \\
1 & 0 
\end{pmatrix}
\begin{bmatrix}
\zs \\
\xs 
\end{bmatrix}\,.
\end{eqnarray}
Consequently, the integral form of the reflected beam is given by
\begin{align}
\label{3dref}
\MoveEqLeft[3]\frac{\psi\re (\xs,y,\zs)}{\exp\left[i\,k\zs\right]} = \inted{k_x}{k_y} R(k_x , k_y)G(k_x,k_y) \nonumber \\
\times{}&\exp\left[i\left( k_x\,\xs + k_y\,y - \frac{k_{\xs}^{\2}+k_y^{\2}}{2k}\,\zs   \right)\right]\,.
\end{align}
In order to obtain an analytical expression for the reflected beam, we consider the first order Taylor expansion
of $R(k_x,k_y)$, 
\begin{equation}
\label{Rapr3d}
R(k_x , k_y ) \approx
 R(0,0)\left( 1+\frac{2\sin\theta}{q\,\cos\varphi}\,k_x\right)\,.
\end{equation}
As done in the previous section, we can rewrite the reflected beam n terms of the incident one as follows
\begin{equation}
\label{refana3d}
\psi\re(\xs,y,\zs) = \mathcal{R}(\xs,\zeta)\psi\inc(\xs,y,\zs)\,,
\end{equation}
where 
\[
\mathcal{R}(\xs,\zeta) = R(0,0)\left[1+ \frac{4\, i\,\xs\sin\theta}
{\w^{\2}\,q\,\cos\varphi}
\frac{1}{1+2i\,\zeta}\right]\,.
\]
To calculate the center of the reflected beam, we have to analyse the term
\begin{eqnarray}
\label{qcvp}
q\,\cos\varphi & = & \sqrt{q^{^2} - k^{^2}\sin^{\2}\theta} \nonumber\\
 & = & k\,\sqrt{(E-V_0)/E - \sin^{\2}\theta}
\end{eqnarray}
and distinguish between real and imaginary values. Then, the analogy between Eqs.\,(\ref{fr}) and (\ref{Rapr3d})
will allow to find the shifts through a simple translation from the results obtained in  the previous section.

\subsection{Incidence with  $E>V_0$}
In this case, it is possible to introduce a \textit{critical} angle
\begin{equation}
\label{critangle}
\theta_{_C} = \arcsin \sqrt{\frac{E-V_0}{E}} \,.
\end{equation}
For $\theta<\theta_{_{C}}$, 
\begin{eqnarray}
\xs^{^\mathrm{centre}}(\theta<\theta_{_{C}}) & = &
\frac{4\,\sin\theta}{k\,\sqrt{\sin^{\2}\theta_{_{C}}\,-\,\sin^{\2}\theta}}\,\,\zeta\\
 & = & \frac{4\,\sin\theta}{(k\,\w)^{^{2}}\,\sqrt{\sin^{\2}\theta_{_{C}}\,-\,\sin^{\2}\theta}}\,\,z\,.
 \nonumber
 \end{eqnarray}
This implies a deviation from the reflection law of geometric optics,
\begin{equation}
\theta_{_\mathrm{REF}}\,\, =\, \theta \,+\, \frac{4\,\sin\theta}{(k\,\w)^{^{2}}\,\sqrt{\sin^{\2}\theta_{_{C}}\,-\,\sin^{\2}\theta}}\, .
\end{equation}
The  velocity change of the reflected particle,  found in the previous section, is now replaced by the angular deviation of the reflected beam.

For  $\theta > \theta_C$, 
\begin{equation}
\xs^{^\mathrm{centre}}(\theta>\theta_{_{C}})\,=\,\frac{2\,\sin\theta}{k\,\sqrt{\sin^{\2}\theta - \sin^{\2}\theta_{_C}}}\,.
\end{equation}
This represents the Goos H\"anchen shift  for the electron and represents the counterpart of the delay time discussed before.

As done in the previous section, by expanding the reflection coefficient around  $\theta_{_C}$,
\begin{equation}
R(k_x,k_y) \approx 1 - 2\,\sqrt{\frac{2\,k_x}{k}\displaystyle\tan\theta_{_C}}\,,
\end{equation}
we can find the mean value fo the shift for critical incidence, $\theta=\theta_{_{C}}$,
 \begin{equation}
 \left\langle \frac{\xs}{\w} \right\rangle
 \,= \, \left( \sqrt{\tan\theta_{c}} \right)\frac{2^{^{\5/\4}}\,\Gamma\left(\displaystyle \frac{5}{4}\right)}{\sqrt{\pi\,k\,\w}}\,\,(\,2\,\zeta\,+\,1\,)\,\,. 
\end{equation}

In figure 3, we plot the angular deviations and the Goos-H\"anchen shift of electron for $k\,\w=500$ (which determins the critical regions), different axial distance $\zeta=0.5,\,1.0,\,2.0$ and two critical angles, i.e. $\arctan 1$ and $\arctan 2$. As expected the difference between mean value and maximum calculations are seen in the critical zone. The analytical predictions (dots) show an excellent agreement with the numerical calculations (continuous lines). For the mean value calculation at the critical angle, we find an amplification between the plots in the panels (a) and (c) given by $\sqrt{2}$. The phenomena show in Figure 3 are very similar to the well-know optical phenomena of laser transmitted through dielectric blocks \cite{DeL19}.

\subsection{Incidence $E<V_0$}

For incidence of a particle with  energy lower than the potential one,  independently of the incidence angle
Eq.\,(\ref{qcvp}) is always imaginary. This implies a shift given by
\begin{equation}
\xs^{^\mathrm{centre}}(E<V_{\0})\,=\,
\frac{\displaystyle 2\sin\theta}{\displaystyle k\,\sqrt{\frac{\displaystyle V_0-E}{\displaystyle E}\,+\,\sin^{\2}\theta }}\,.
\end{equation}
Analytical predictions (dots) and numerical results (continuous lines) appear in the plots of Figure 4, for two values of $k\,\w$, i.e. 500 and 1000, and three values of $\sqrt{V_{\0}/E}$, i.e. 1.10, 1.30, and 1.70. In the next section, we shall see as this shift find its natural  optical counterpart in the reflection by dielectric with imaginary refractive index.

\section{Optical analogy}
Let us consider, an optical beam reflected at a dielectric/air interface. If $\theta$ and $\varphi$ are the incidence and refraction angles and $n$ the refractive index od the dielectric,
\begin{equation}
    n\,\sin{\theta}\,=\,\sin{\varphi}\,.
\end{equation}
Total reflection is seen when the incidence angle is greater than
\begin{equation}
    \theta_{_\mathrm{C}}\,=\,\arcsin (1/n)\,.
\end{equation}
From  Eq.\eqref{critangle}, we immediately see that a the quantum mechanical results can be translated in its optical counterpart by using the following translation rule
\[ E/(E-V_{\0}) \,\,\,\leftrightarrow\,\,\, n^{\2}\,.   \]
For $E<V_{\0}$ we find imaginary refractive index.

In the case of optical beams, the reflection coefficients used to obtain an analytical expression for the reflected beam are given by
\begin{equation}
R^{^\mathrm{[\sigma]}}\left(k_x,k_y\right)\,\approx\,
R^{^\mathrm{[\sigma]}}\left(0,0\right)\left(
1+\alpha^{^\mathrm{[\sigma]}}\,\frac{k_x}{k}\right)\,,
\end{equation}
where $\sigma$=TE, TM and 
\begin{equation}
\begin{Bmatrix}
\alpha^{^\mathrm{[TE]}} \\
\\
\alpha^{^\mathrm{[TM]}}
\end{Bmatrix}
=
2\,\tan\varphi 
\begin{Bmatrix}
\displaystyle
1 \\
\\
\displaystyle\frac{1}
{n^{2}\sin^{2}\theta - \cos^{2}\theta}
\end{Bmatrix}\,.
\end{equation}

\section{Conclusions}
In this paper, by using a Taylor expansion of the reflection coefficient we obtained a closed expression
for the shift of the center of the beam reflected by   one-dimensional potential in quantum mechanics. Two kinds of  displacements were  found. The time-dependent shift is a clear  evidence of the change in the velocity of the reflected particle  not only in the direction but also in its modulus. This phenomenon seems to be never predicted
before. The time independent shift is related to the well-know delay time.

By extending our analysis to three dimensional problems, we find two phenomena strictly related to the previous ones:
angular deviations and Goos-H\"anchen shift for the electron. By a simple translation rule, we can thus reproduce in an optical laboratory quantum mechanical experiments, in particular indirect measurements of delay times.

\subsection*{Acknowledgements}

One of the authors (S.D.L.) thanks the CNPq (grant 2018/303911) and
Fapesp (grant 2019/06382-9) for financial support. The authors are also  grateful 
to A. Alessandrelli, L. Maggio, A. Stefano and, in particular, to  D. Lotito   for their scientific comments and suggestions  and to Profs. G. Co$^\prime$, L. Girlanda, and A. Nucita
for their help in consolidating the research \textit{BRIT} project of international collaboration between  the State University of Campinas (Brazil) and the Salento University of Lecce (Italy). The authors also thanks an anonymous referee for his suggestions and recommendations.


\figureA
\figureB
\figureC
\figureD

\end{document}